\documentclass{ws-ijmpa}

\begin{document}

%
\catchline{}{}{}{}{}
%

\title{STRUCTURE OF $\Lambda^*$(1405) AND THE $\Lambda^*$-MESON-BARYON COUPLING CONSTANTS}

\author{C. S. AN, B. SAGHAI}

\address{CEA, Centre de Saclay, IRFU/Service de Physique Nucléaire, F-91191 Gif-sur-Yvette, France\\
chunsheng.an@cea.fr ; bijan.saghai@cea.fr}

\maketitle

\pub{Received (Day Month Year)}{Revised (Day Month Year)}

\begin{abstract}
Within an extended chiral constituent quark model, three- and five-quark structure of the $S_{01}$
resonance $\Lambda(1405)$ is investigated with respect to the coupling constants 
$g^2_{\Lambda^* \pi \Sigma}$ and $g^2_{\Lambda^* \bar{K} N}$.  
Our findings corroborate with
about 50\%
of five-quark admixture in the $\Lambda(1405)$ needed in reproducing the strong decay width,
$\Gamma_{\Lambda(1405)\to (\Sigma \pi)^\circ}$.

\end{abstract}

\section{Introduction and Theoretical Frame}  

The nature of the $\Lambda^*$(1405)-resonance, investigated since half a century, still bears
puzzling features. Its well established couplings to $\bar K N$ and $\pi \Sigma$ states have
offered guidance to various theoretical approaches in improving our understanding of it.
 
Recent achievements\cite{5q} in describing non-strange baryons such as the nucleon, $\Delta$-, $P_{11}$(1440) and 
$S_{11}$(1535)-resonances as a superposition of three- and five-quark states bring in new insights
into the structure of baryons.

In a recent work\cite{An} a chiral constituent quark model approach was extended to the strangeness sector,
studying the $\Lambda^*$(1405) in a truncated Fock space, which includes three- and five-quark components, as well as configuration mixings among them,
namely, $qqq \leftrightarrow qqqq \bar{q}$ transitions. 
That formalism allowed us to calculate the helicity amplitudes for the electromagnetic decays
($\Lambda^* \to \Lambda(1116)\gamma$, $\Sigma(1194)\gamma$), and transition amplitudes for
strong decays ($\Lambda^*\to\Sigma(1194)\pi$, $ K^{-}p$), as well as the relevant
decay widths, namely, $\Gamma_{\Lambda^*\rightarrow \Lambda(1116) \gamma}$, 
$\Gamma_{\Lambda^* \to \Sigma(1193)\gamma}$,
and $\Gamma_{\Lambda(1405) \to \Sigma(1194) \pi}$.
The only available experimental value\cite{PDG}, for the strong decay width 
$\Gamma_{\Lambda^*\to (\Sigma \pi)^\circ}$, was well reproduced with about 50\% 
of five-quark admixture in $\Lambda^*$.

In this contribution we concentrate on the coupling constants $g_{\Lambda^* \bar K N}$ 
and $g_{\Lambda^* \pi \Sigma}$ allowing us to put further constraints on the percentage
of the five-quark component within the $\Lambda^*$. The starting point is 
the hadronic level Lagrangian for the
$\Lambda(1405) B M$ coupling, with $B \equiv \Sigma,~N$ and $M \equiv \pi,~K$
\begin{equation}
\mathcal{L}_{\Lambda(1405)BM}=i\frac{f_{\Lambda(1405)BM}}{m_{M}}\bar{\psi}_{B}\gamma_{\mu}
\partial^{\mu}\phi_{M}X_{M}\psi_{\Lambda(1405)}+h.c.,
\end{equation}
where the transition coupling amplitude reads
\begin{equation}
\frac{f_{\Lambda(1405)BM}}{m_{M}}=\frac{\langle[\hat{T}^{M}_{d}+\hat{T}^{M}_{35}+\hat{T}^{M}_{53}]\rangle}
{m_{\Lambda(1405)}-m_B},
\end{equation}
with the diagonal ($\hat{T}^{M}_{d}$) and non diagonal ($\hat{T}^{M}_{53}$ and $\hat{T}^{M}_{35}$)
transition amplitudes calculated within the nonrelativistic chiral constituent quark model\cite{An}.
The $\Lambda ^*$-Baryon-Meson coupling constant is given by
\begin{equation}
g_{\Lambda ^*BM} =  \frac{m_B-m_{\Lambda^*}}{m_M} f_{\Lambda^*BM}.
\end{equation}
%

\section{Results and Discussion}

The coupling constants $g_{\Lambda^* \bar K N}$ and $g_{\Lambda^* \pi \Sigma}$, 
as well as the ratio $R=g_{\Lambda^* \bar K N}/g_{\Lambda^* \pi \Sigma}$ have been 
investigated both experimentally\cite{Exp} and within various theoretical 
approaches\cite{Theo,Kim,Weil,G-M,Dalitz,Martin,Oneda}, but
none of them relying on the internal quark structure of the $\Lambda^*$-resonance. 
Here we report on the results obtained within our chiral constituent quark approach, and
investigate the dependence of the coupling constants on the percentage of genuine five-quark 
admixture in the  $\Lambda^*$ wave function. 
 
 Values for those entities, extracted through a T-matrix effective-range expansion\cite{Kim} are
\vspace{-3pt}
$$g^2_{\Lambda^* \pi \Sigma}/4 \pi=0.047 \pm 0.007~;~g^2_{\Lambda^* \bar{K} N}/4 \pi=0.32 \pm 0.02~;~R={\frac{g^2_{\Lambda^* \bar{K} N}}{g^2_{\Lambda^* \pi \Sigma}}=6.8\pm1.0}~.$$
%

Several authors report results for the ratio $R$ and not always for 
individual coupling constants.
The ratio R, given above, comes out to be about one order of magnitude larger than its value (2/3) 
if it were a pure $SU(3)$  singlet. 
It is also significantly different from values obtained by various approaches, such as
current algebra\cite{Weil,G-M}: 3.2, 
potential models\cite{Weil,Dalitz}: 4.8, 
dispersion relations\cite{Martin}: 4.0,
or still asymptotic $SU(3)$ symmetry approach\cite{Oneda}: 4.8.

\begin{figure}[t!]
\begin{center}
\parbox[c]{0.46\textwidth}{
\centering
\includegraphics[width=0.5\textwidth]{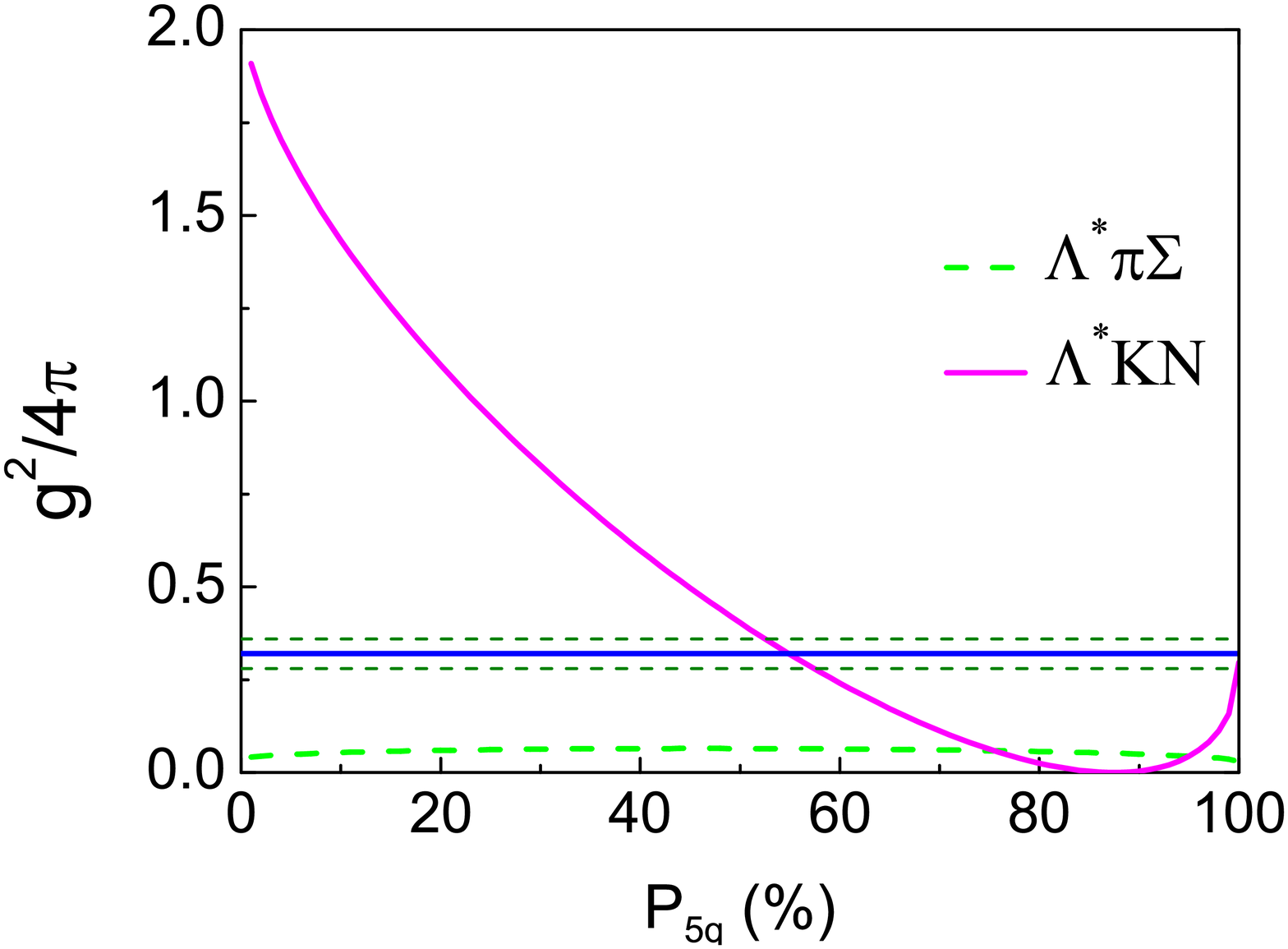}
}
\parbox[c]{0.5\textwidth}{
\centering
\includegraphics[width=0.5\textwidth]{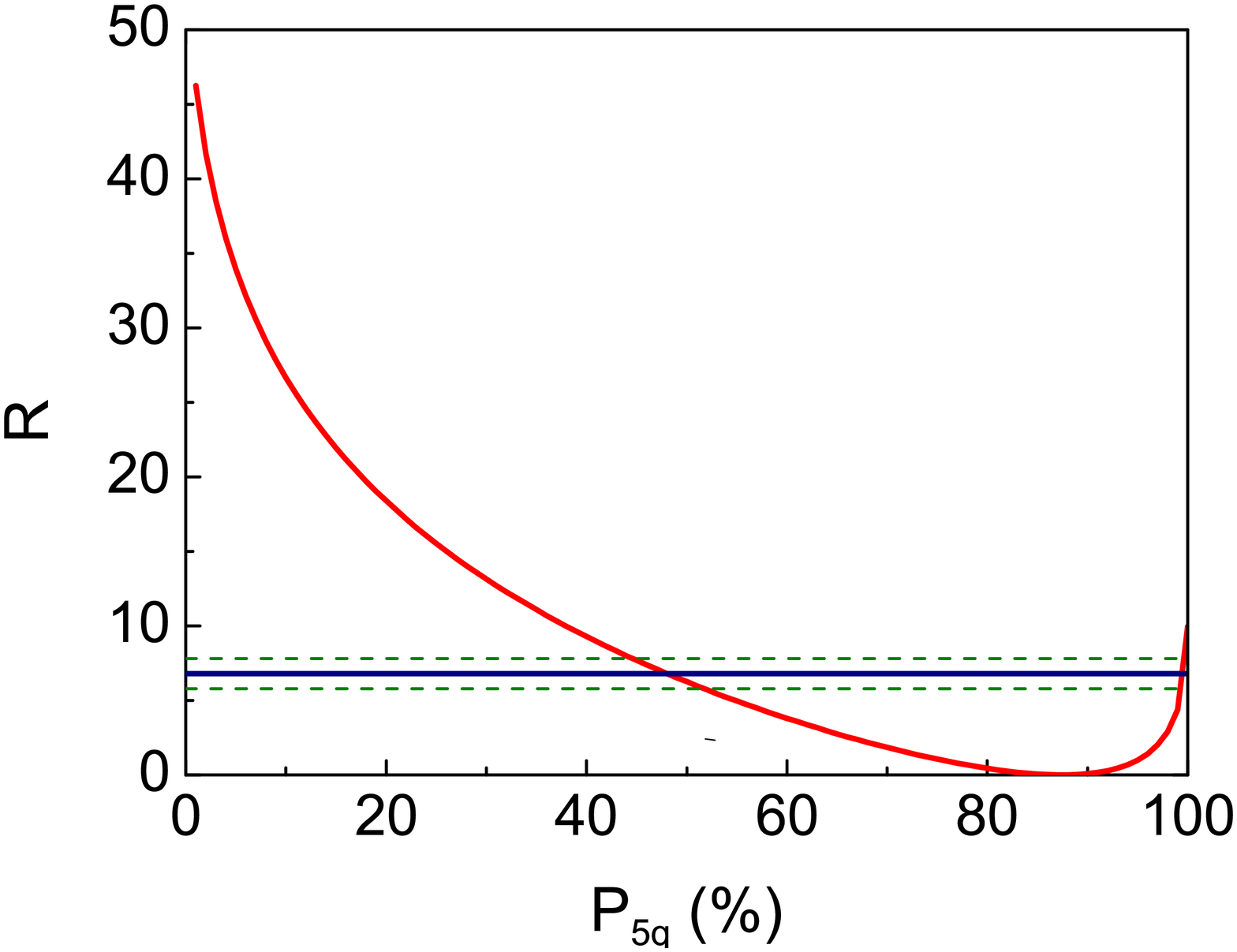}
}
\vspace*{8pt}
\caption{{\bf Left}: Coupling constants $g_{\Lambda^* \bar K N}$ 
and $g_{\Lambda^* \pi \Sigma}$ as a function of the five-quark component ($P_{5q}$) in $\Lambda^*(1405)$
{\bf Right}: ratio $R=g_{\Lambda^* \bar K N}/g_{\Lambda^* \pi \Sigma}$ as a function of $P_{5q}$.
For explanations of horizontal lines see the text.}
\label{fig1}
\end{center}
\end{figure}

In Fig.~\ref{fig1} our results are shown.
In the Left panel coupling constants as a function of five-quark component percentage ($P_{5q}$) 
in the $\Lambda^*$ wave function are depicted. 
As known from other sources, the $\bar{K} N$ coupling to $\Lambda^*$ is (much) larger than coupling to
$\pi \Sigma$. The latter, within our approach, shows no significant sensitivity to $P_{5q}$.
Actually, the predicted value for $g^2_{\Lambda^* \pi \Sigma}$ starts and ends at 0.031,
after having gone through a maximum around 0.065 at $P_{5q}\approx$46\%. 
This smooth dependence on $P_{5q}$ does not
impose significant constraints on the $P_{5q}$ range.
On the contrary, the $g^2_{\Lambda^* \bar{K} N}$ varies, at least up to $P_{5q} \lesssim$ 60\%, rather drastically.
The horizontal line corresponds to the central value in $g^2_{\Lambda^* \bar{K} N}/4 \pi=0.32 \pm 0.02$
and dotted lines to $\pm \sigma$,
intercepting the prediction curve at $P_{5q}=(55 \pm 1$)\%. 

In the Right panel, Fig.~\ref{fig1}, our results for the ratio $R$ as a function of $P_{5q}$ are shown.
The horizontal lines correspond to $R=6.8 \pm 1.0$. 
We notice that the smooth variation of $g^2_{\Lambda^* \pi \Sigma}$
affects nevertheless the shape of $R$ and the intersection values. 
Actually, from that figure we deduce
$P_{5q}=(48 \pm 3)$\%, in agreement with the $P_{5q} \approx 50 $\% found to reproduce the strong 
decay width  $\Gamma_{\Lambda(1405)\to (\Sigma \pi)^\circ}=50\pm 2$ MeV.

In conclusion, our recent\cite{An} and present studies strongly suggest an admixture of five-quark components
in $\Lambda^*$ at the level of $P_{5q} \approx 50 $\%. 
Extensive ongoing theoretical investigations (see e.g. Refs.\cite{An,Jido} and references therein) 
will greatly benefit
from current experimental programmes on the $K^-$-nucleon interactions in $DA \Phi NE$\cite{DAFNE} 
and electromagnetic production of $\Lambda^*(1405)$ in JLab\cite{JLab}.

\begin{thebibliography}{00}    
%
\bibitem{5q} Q.~B.~Li and D.~O.~Riska, {\bf Phys.\ Rev.\  C 73}, 035201 (2006);
{\it ibid} {\bf C 74}, 015202 (2006); 
{\bf Nucl.\ Phys.\  A 766}, 172 (2006); 
B.~Julia-Diaz and D.~O.~Riska, {\it ibid} {\bf A 780}, 175 (2006);
C.~S.~An and B.~S.~Zou, {\bf Eur.\ Phys.\ J.\  A 39}, 195 (2009).
%
\bibitem{An} C.~S.~An, B.~Saghai, S.~G.~Yuan and J.~He,
  {\bf Phys.\ Rev.\  C 81}, 045203 (2010).
%
\bibitem{PDG} C.~Amsler {\it et al.}  [Particle Data Group],
  {\bf Phys.\ Lett.\  B 667}, 1 (2008).
%
\bibitem{Exp}
%
R. D. Tripp {\it et al.}, {\bf Phys.\ Rev.\ Lett.  21}, 1721 (1968);
  O.~Braun {\it et al.},
  {\bf Nucl.\ Phys.\  B 129}, 1 (1977).
%
\bibitem{Theo}
J.~Soln, {\bf Phys.\ Rev.\ D 2}, 2404 (1970);
D. L. Katyal and A. N. Mitra, {\it ibid} {\bf D 1}, 338 (1970);
G. Rajasekaran, {\it ibid} {\bf D 5}, 610 (1972);
G.~C.~Oades and G.~Rasche, {\bf Nuovo Cim.\  A  42}, 462 (1977).
%
\bibitem{Kim}
J.~K.~Kim and F.~Von Hippel, {\bf Phys.\ Rev.\ 184}, 1961 (1969).
%
\bibitem{Weil}
C. Weil, {\bf Phys.\ Rev.\ 161}, 1617 (1967).
%
\bibitem{G-M}
M. Gell-Mann R. J. Ookes and B. Renner, {\bf Phys.\ Rev.\ 175}, 2195 (1968).
%
\bibitem{Dalitz}
R.~H.~Dalitz, T.~C.~Wong and G.~Rajasekaran, {\bf Phys.\ Rev.\  153}, 1617 (1967).
%
\bibitem{Martin}
A. D. Martin, {\bf Phys.\ Lett.\ B 65}, 346 (1976).
%
\bibitem{Oneda} 
S.~Oneda and S.~Matsuda, {\bf Phys.\ Rev.\  D 2}, 887 (1970).
%
\bibitem{Jido}
  D.~Jido, J.~A.~Oller, E.~Oset, A.~Ramos and U.~G.~Meissner,
 {\bf  Nucl.\ Phys.\  A 725}, 181 (2003);
 D.~Jido, T.~Sekihara, Y.~Ikeda, T.~Hyodo, Y.~Kanada-En'yo and E.~Oset,
  {\it ibid} {\bf  A 835}, 59 (2010).
%
\bibitem{DAFNE}
  M.~Cargnelli {\it et al.},
  {\bf  Nucl.\ Phys.\  A 835}, 27 (2010).
%
\bibitem{JLab}
  K.~Moriya and R.~Schumacher  [CLAS Collaboration],
  {\bf  Nucl.\ Phys.\  A  835}, 325 (2010).
%
\end{thebibliography}
\end{document}